\documentclass[twocolumn]{aastex631}
\usepackage{graphicx}	
\usepackage{epstopdf}   
\usepackage{amsmath}	
\usepackage{amssymb}	
\usepackage[T1]{fontenc}   
\usepackage{ae,aecompl}
\usepackage{float}    
\usepackage{textcomp}   
\usepackage{color}
\usepackage{yhmath}   

\usepackage{lipsum}  
\usepackage{orcidlink}   
\shorttitle{RVM and RFM with multipole magnetic field}
\shortauthors{J. L. Qiu, H. Tong, H. G. Wang}
\graphicspath{{./}{figures/}}

\begin{document}
			
\title{Rotating vector model and radius-to-frequency mapping in the presence of multipole magnetic field}

\correspondingauthor{H. Tong}
\email{tonghao@gzhu.edu.cn}
			
\author{J. L. Qiu, H. Tong~\orcidlink{0000-0001-7120-4076}, H. G. Wang}
\affiliation{School of Physics and Materials Science, Guangzhou University, Guangzhou 510006, China}
			
			
			
			
			
			
			
			
			

\begin{abstract}
	The rotating vector model and radius-to-frequency mapping in the presence of multipole magnetic field in pulsars and magnetars are considered. An axisymmetric potential field is assumed. It is found that: (1) The radiation beam in the case of multipole field is wider than the dipole case. This may account the increasing pulse width at higher frequency of pulsars (anti-radius-to-frequency mapping). (2) The expression for the polarization position angle is unchanged. Only the inclination angle $\alpha$ and phase constant $\phi_0$ will change. The angle between the rotational axis and line of sight, and the position angle constant $\psi_0$ will not change. When fitting the varying position angle of magnetars, these constraints should be considered.  The appearance and disappearance of multipole field may account for the changing slope of position angle in the radio emitting magnetar Swift J1818.0-1607. Similar but more active process in magnetar magnetospheres may account for the diverse position angle in fast radius bursts.
\end{abstract}



\section{Introduction}

The multiwave emission of pulsars originates in their magnetospheres (Goldreich \& Julian 1969; Ruderman \& Sutherland 1975; Cheng et al. 1986; Du et al. 2010). The large-scale magnetic field in the magnetosphere is assumed to be dominated by the dipole component, with possible contribution of various multipole magnetic field (Bilous et al. 2019).

The rotating vector model (RVM) is a model for pulsar radio emissions (Radhakrishnan \& Cooke 1969). It combines the geometry of the pulsar's magnetospheric radiation with the observer’s line of sight, and can account for the polarized position angle swing.

Most radio pulsars are found to have narrow average profiles. For instance, in Fig.4 of Johnston \& Kramer (2019), the pulse profiles of radio pulsars only account for a small part of the whole phase cycle. In addition, it can be seen from the duty cycle of pulses shown in Table B.2 of Pilia et al.(2016), as well as Fig. 3-6 of Rankin (1993), that most of the profiles comprise only about 10\% of the entire period, or even less. 
Due to the narrow profile of most radio pulsars and the small amount of observed data, the position angle of the linearly polarized radiation follows linear trend within the on-pulse window. In Radhakrishnan \& Cooke (1969), when the line of sight passes through the radiation beam, the angle between the line of sight and the magnetic field line in radiation beam changes with time, which is summarized as RVM. The position angle (which is written as $\psi$ in this article) is a function of pulse longitude (Johnston \& Kramer 2019):
\begin{equation} \label{RVM_PA}
	\psi=\psi_{\rm 0}+\arctan\frac{\sin{\alpha}\sin{(\phi-\phi_{\rm 0})}}{\cos{\alpha}\sin{\zeta}-\sin{\alpha}\cos{\zeta}\cos(\phi-\phi_{\rm 0})},
\end{equation}
where $\zeta=\alpha+\beta$, $\alpha$ is the magnetic inclination angle, and $\beta$ is the impact angle (represents the angle of closest approach of the line of sight to the magnetic axis), and $\phi_{\rm 0}$ is the pulse longitude when $\psi=\psi_{\rm 0}$. If RVM is used to fit pulse polarization, degeneracy of some parameters such as $\alpha$ and $\beta$ will occur. Equation (\ref{RVM_PA}) and the maximum slope value of position angle curve
\begin{equation}
	(\frac{d\psi}{d\phi})_{\rm max}=\frac{\sin{\alpha}}{\sin{\beta}}
\end{equation}
are used to fit the position angle curve to acquire $\alpha$ and $\beta$ within a range, such as the Fig. 1 of Johnston et al. (2023).
Therefore, application of RVM to pulsar observations can only constrict the inclination angle and viewing angle of the pulsars (Lyne \& Manchester 1988; Rankin 1993; Manchester et al. 1998; Everett \& Weisberg 2001; Johnston \& Weisberg 2006; Johnston \& Kramer 2019).

In the original version of RVM a large scale dipole magnetic field is assumed (Radhakrishnan \& Cooke 1969; Komesaroff 1970). The effect of rotation (Blaskiewicz et al. 1991; Wang et al. 2012), polar cap currents (Hibschman \& Arons 2001) will result in modifications to RVM. In the case of magnetars, the magnetic field may be a twisted dipolar (Thompson et al. 2002; Tong 2019). Then the RVM for magnetars will also change to some degree (Tong et al. 2021).

Multipole fields may be present in pulsars and magnetars. In Arumugasamy et al. (2018), from the X-ray spectrum of PSR J0659+1414, the required magnetic field for the absorption producing at the energy $E \approx 0.54$ keV is 24 times larger than the canonical dipole field of PSR J0659+1414. As stated by Bilous et al. (2019), the recently NICER observations of thermal X-ray pulsations from the surface of PSR J0030+0451 indicates the non-antipodal hot emitting regions on the pulsar’s  surface. Both works suggest the presence of a multipolar field near pulsar's surface.
In the existence of multipole magnetic field in pulsars and magnetars (Thompson et al. 2002; Pavan et al. 2009; Beloborodov 2009; Tong 2019; Bilous et al. 2019), the rotating vector model may also change.

	Two observations inspired us to consider RVM in the presence of multipole magnetic field.
\begin{enumerate}
  \item The anti-radius-to-frequency mapping of some pulsars. The pulses in different frequencies are generated at different radiation heights.
  Radius-to-frequency mapping (abbreviated as RFM) is an observed phenomenon that depicts the narrowing width of radio radiation profile of a pulsar with the increasing frequency, and the anti-RFM is the opposite, as seen in some of pulsars observed in Chen \& Wang (2014), Xu et al. (2021) and Posselt et al. (2021).
  Theoretically, the dipole magnetic field is a relatively credible magnetosphere model, and the radio radiation originate from the open field line region (that is, at the pulsar’s magnetic pole). It is considered in this paper that the radio emission of pulsar belongs to the narrowband radiation, that is, the radiation in a specific frequency is generated at a emission height. For a dipole field, the pulse width is expected to be narrower at higher frequency (Ruderman \& Sutherland 1975; Cordes 1978; Wang et al. 2013). Many of the non-recycled pulsars exhibit this kind of behaviour, i. e. RFM (Chen \& Wang 2014; Posselt et al. 2021). If RFM is due to a dipole geometry, then the anti-RFM may imply that the magnetic field is no longer a pure dipole, i.e. multipole magnetic field may be there.
  
  \item As it is observed, the position angle of the linearly polarized radiation component is set by the local plane of curvature of the field line (Lyne \& Graham-Smith 2012). In RVM, the slope of the polarization position angle is determined by the inclination angle and impact angle. It is fixed by the magnetic axis and line of sight. Therefore, a change of the slope of position angle is unimaginable for a dipole geometry, and there is some  possibility that this is caused by a twisted dipole field with one more toroidal component (Tong et al. 2021). However, a change slope of position angle is indeed observed in the radiating magnetar Swift J1818.0-1607 (Lower et al. 2021). In this paper, we suspect that the appearance and disappearance of multipole field with time may give rise to the change of position angle of Swift J1818.0-1607.

      Fast radio bursts (FRBs) are the bursts with extremely short duration (on the order of milliseconds), extremely high energy (up to $\sim 10^{41}$ erg) and large dispersion measures in radio band. A more diverse position angle exists in the case of FRB (Luo et al. 2020). A similar but more active magnetosphere of magnetars may account for observational properties of FRBs. The greatly high brightness temperature (Lorimer et al. 2007) means the extremely coherent emission mechanism of FRBs. Although the radiation mechanism of FRBs is still not understood, magnetosphere origin of FRB is supported by more and more evidence, such as the diversity of period-folded position angle features of there bursts (Luo et al. 2020), and constraints on high radio radiation efficiency and other parameters (Zhang 2022). According to Lu \& Kumar (2018) and Bochenek et al. (2020), the magnetosphere of magnetar produces its radio emission of FRB. We will focus on the observations of magnetars, since more quantitative information can be obtained there, such as the magnetic inclination angle $\alpha$, position angle $\psi$, and width of each components in pulse profile, which are related to magnetosphere structure, together with parameters like spertral index, which can reveal the radiation mechanism (Lower et al. 2021).
\end{enumerate}

The paper is organized as follows. The description of the multipole field is presented in Section 2. The relation between the emission point and the line of sight is explored in Section 3. The modifications of RVM and RFM in the presence of multipole field are calculated in Section 4 and 5, respectively. Discussion and conclusion are given in Section 6 and Section 7, respectively.

\section{Description of the multipole magnetic field}
\label{section_multipole magnetic field}

The original rotating vector model was developed for the inclined dipolar magnetic field (Radhakrishnan \& Cooke 1969; Komesaroff 1970). For a twisted dipole magnetic field of magnetars, the corresponding modification presented in Tong et al. (2021) shows that there is an additional toroidal component on the basis of a pure dipole field, and then the position angle can be written as $\psi=\psi(\rm dipole)+\bigtriangleup \psi_{\rm twisted}.$

A general formula for estimating the correction of position angle due to a toroidal magnetic field component is also presented there (equation (30) in Tong et al. 2021). Therefore, here we can focus on the potential field and simplify our description of multipole magnetic field.

The magnetosphere of pulsars has been modeled by many researches, such as Roberts (1979), Bonazzola et al. (2015) and Petri (2015). It is generally believed that the radio radiation of pulsar originates from their magnetosphere. Thus the various observed properties of radio emission from pulsars is helpful to understand the structure of pulsars’ magnetosphere and the radiation mechanism involved. The dipole magnetic field is the most basic assumption for pulsar’s magnetosphere, and the radiation mechanism of a pulsar involves assumption of magnetospheric geometry, modeling of gaps, assumption about distribution of particles and so on. They can be combined to interpret the observations, such as RFM (Komesaroff 1970), nulling (Zhang et al. 1997), and subpulse drifting (Ruderman \& Sutherland 1975). Based on previous works, we simplify the magnetosphere model to a certain extent in this paper.

The simple axisymmetric case is considered, corresponding to special case in Bonazzola et al. (2015) and Petri et al. (2015). The multipole magnetic field in vacuum is also used in this paper, assuming force-free condition of the pulsar and magnetar's magnetosphere (Wolfson 1995; Thompson et al. 2002; Pavan et al. 2009; Beloborodov 2009; Fujisawa \& Kisaka 2014; Akg\"{u}n et al. 2016; Kojima 2017; Tong 2019). In the special potential field case, the magnetic field can be expressed as the gradient of the scalar potential (Wiegelmann \& Sakurai 2012)
\begin{equation}
  {\bf{B}} = -\nabla \phi.
\end{equation}
The scalar potential satisfies the Laplace equation
\begin{equation}
  \nabla^2 \phi =0.
\end{equation}
In the axisymmetric case, the general solution of scalar potential is (in the magnetic frame)
\begin{equation}
  \phi(r,\theta)= \sum_{l} B_l r^{-l-1} P_l(\cos\theta),
\end{equation}
where $B_l$ are numerical constants, $P_l (x)$ are Legendre polynomials. Here only the exterior solution to the Laplace equation is considered. For $l=0$, it corresponds to the Coulomb field case; meanwhile, it means a magnetic monopole for the magnetic field case, and does not make sense physically. For $l=1$, it is the dipole field: $B_1 \frac{1}{r^2} \cos\theta$. For $l=2$, it is the quadrupole field: $B_2 \frac{1}{r^3} (3\cos^2\theta-1)/2$. For $l=3$, it is the octupole field case: $B_3 \frac{1}{r^4} (5\cos^3\theta - 3\cos\theta)/2$, etc. The expression for magnetic field can be obtained by the gradient of the scalar potential (noting that we are using spherical coordinate in the magnetic frame). For example, the dipole magnetic field is:
\begin{eqnarray}
  {\bf{B}}_{\rm dip} &=& -\nabla \left( \frac{B_1}{r^2} \cos\theta \right)\\
  &=& -\left( \hat{r} \frac{\partial}{\partial r} + \hat{\theta} \frac{1}{r} \frac{\partial}{\partial \theta} \right) \left( \frac{B_1}{r^2} \cos\theta \right)\\
  &=& \frac{B_1}{r^3} 2(\cos\theta \hat{r} + \frac{1}{2}\sin\theta \hat{\theta}).
\end{eqnarray}
From this expression it can be seen that the parameter $B_1$ is the dipole moment of magnetic field. The quadrupole field is:
\begin{equation}\label{eqn_quadrupole}
  {\bf B}_{\rm quad} = \frac{B_2}{r^4} \ 3 \ \left( \frac{3\cos^2\theta-1}{2} \hat{r} + \sin\theta \cos\theta \hat{\theta} \right).
\end{equation}
As for the general multipole field order $l$, the expression for magnetic field is:
\begin{equation}\label{eqn_multipole}
  {\bf B}_{l} = \frac{B_l}{r^{l+2}} \ (l+1) \ \left[ P_l(\cos\theta) \hat{r} + \frac{P_l^{\prime}(\cos\theta)}{l+1}\sin\theta \hat{\theta} \right].
\end{equation}

The magnetic field line is defined as a line with tangent at every point parallel to the local magnetic field. The potential field has only poloidal component, but no toroidal component. Therefore, the magnetic field line is (in polar spherical coordinate): $dr/B_r = r d\theta/B_{\theta}$, or $dr/d\theta = r (B_r/B_{\theta})$. For a dipole field, the equation can be integrated and the equation for the field line is: $r=r_{\rm e} \sin^2\theta$ ($r_{\rm e}$ is the maximal radial extension of the field line). For a quadrupole field, the field line obeys the equation: $dr/d\theta = (1/2) r (3\cos^2\theta-1)/(\sin\theta \cos\theta)$. For the general multipole field with order $l$, the field line is governed by the differential equation:
\begin{equation}
  \frac{dr}{d\theta} = r (l+1) \frac{P_l(\cos\theta)}{P_l^{\prime}(\cos\theta)\sin\theta}.
\end{equation}
A more convenient way for obtaining the equation of field line is introducing the vector potential and flux function. The potential field is a special case of the Grad-Shafranov equation for axisymmetric force-free equilibria (Wolfson 1995). Constants of $A=r^{-l} f(x)$ correspond to the magnetic field lines, where $f(x)=\sqrt{1-x^2} P_l^1(x)$ ($x=\cos\theta$, $P_l^1$ is the associated Legendre functions\footnote{The symbol $l$ here corresponds to the symbol $n$ in Wolfson (1995)}.). The polar cap angles of pulsars are different in flat and curved space-time (Gonthiner \& Harding 1994), but non-recycled pulsars are considered in this paper, so the influence of general relativity on the shape and size of the polar caps is ignored. In addition, considering the large distance scale, it is assumed that the magnetic field lines are approximately the same in flat and curved space-time.

\begin{figure}
	\centering
	\includegraphics[width=0.47\textwidth]{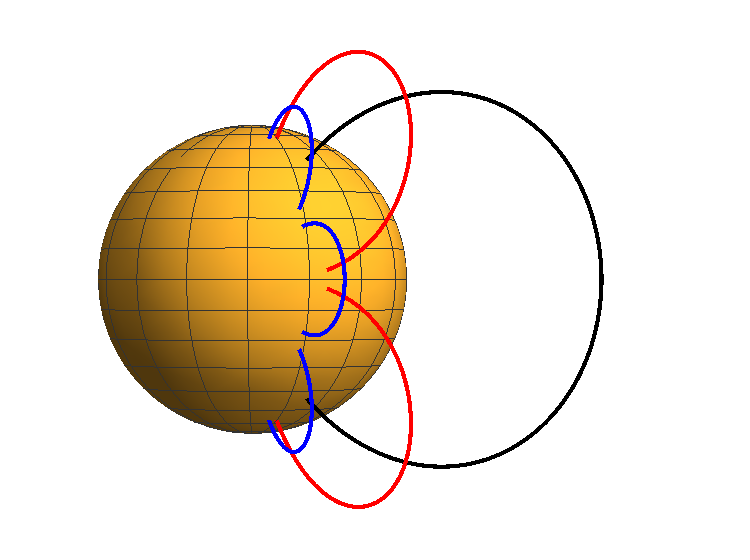}
	\caption{Dipole (black), quadrupole (red), and octupole (blue) magnetic field lines in three-dimensional of a neutron star. Each field line is plotted according to its equation. The radial extent is arbitrary and is for schematic use only.}
	\label{3Dfig_gmultipole}
\end{figure}

The three-dimensional dipole, quadrupole and octupole magnetic fields in a neutron star are shown in Fig. \ref{3Dfig_gmultipole}.

\section{Relation between emission point and line of sight}
\label{section_Relation}

It is generally assumed that the tangent vector at the emission point is parallel to the line of sight (Hibschman \& Arons 2001). Given the magnetic field line geometry, the relation between emission point and line of sight can be calculated. The calculations for a dipole field can be found in Appendix C in Tong et al. (2001). The procedure for a multipole field is similar. Below we give an example of such calculation for the quadrupole magnetic field.

Here the magnetic frame is considered as a spherical coordinate system with the magnetic field center as the origin and the magnetic axis as the z axis. In the magnetic frame, the line of sight lies along the direction defined by $(\theta_{\rm obs},\phi_{\rm obs})$. In Cartesian coordinate system, the unit vector along the line of sight is:
\begin{equation}
  \hat{l} = \sin\theta_{\rm obs} \cos\phi_{\rm obs} \hat{x} + \sin\theta_{\rm obs}\sin\phi_{\rm obs} \hat{y} + \cos\theta_{\rm obs} \hat{z}.
\end{equation}
From the expression for quadrupole field (equation (\ref{eqn_quadrupole})), the tangent vector of magnetic field at point $(r,\theta,\phi)$ is:
\begin{equation}
  \hat{t} = \frac{\bf B}{|{\bf B}|} = \frac{1}{N} \left( \frac{3\cos^2\theta - 1}{2} \hat{r} + \sin\theta \cos\theta \hat{\theta} \right),
\end{equation}
where $\rm N=\sqrt{6\cos(2\theta)+\frac{5}{2}[3+\cos(4\theta)]}/2$ is the normalization constant. For a small $\theta$, we use the estimations $\sin\theta \approx \theta$ and $\cos\theta \approx 1$, and then N is $2$ approximately. Transforming tangent vector from spherical to Cartesian coordinate system (see Appendix C in Tong et al. 2021):
\begin{eqnarray}
\nonumber
  \hat{t} &=& \frac{1}{N} ( \frac{5\cos^2\theta - 1}{2} \sin\theta \cos\phi \hat{x} + \frac{5\cos^2\theta -1}{2} \sin\theta \sin\phi \hat{y} \\
  &&+ \frac{5\cos^2\theta -3}{2}\cos\theta \hat{z} ).
\end{eqnarray}
For the radio emission of pulsars and magnetars, it is generally assumed that: $\hat{t} \parallel \hat{l}$ (Ruderman \& Sutherland 1975; Hibschman \& Arons 2001). Then $x,\ y,\ z$-component of the tangent vector are equal to the unit vector along the line of sight:
\begin{eqnarray}\label{eqn_line_of_sight_1}
  \frac{1}{N} \frac{5\cos^2\theta -1}{2} \sin\theta \cos\phi &=& \sin\theta_{\rm obs} \cos\phi_{\rm obs} \\
  \label{eqn_line_of_sight_2}
  \frac{1}{N} \frac{5\cos^2\theta -1}{2} \sin\theta \sin\phi &=& \sin\theta_{\rm obs} \sin\phi_{\rm obs} \\
  \label{eqn_line_of_sight_3}
  \frac{1}{N} \frac{5\cos^2\theta -3}{2} \cos\theta &=& \cos\theta_{\rm obs}.
\end{eqnarray}
Equation (\ref{eqn_line_of_sight_2}) divided by equation (\ref{eqn_line_of_sight_1}) gives: $\tan\phi = \tan\phi_{\rm obs}$. Therefore, $\phi=\phi_{\rm obs}$. For a poloidal field, the line of sight lies at the same meridian plane of the magnetic field line. Then equation (\ref{eqn_line_of_sight_2}) can be simplified to:
\begin{equation}\label{eqn_line_of_sight_2_simplify}
  \frac{1}{N} \frac{5\cos^2\theta -1}{2} \sin\theta = \sin\theta_{\rm obs}.
\end{equation}
Equation (\ref{eqn_line_of_sight_2_simplify}) divided by equation (\ref{eqn_line_of_sight_3}) gives:
\begin{equation}\label{eqn_thetaobs_and_theta}
  \tan\theta_{\rm obs} = \tan\theta \frac{5\cos^2\theta -1}{5\cos^2\theta -3}.
\end{equation}
This is the relation between the emission point at $(r, \ \theta)$ and the size of the emission cone. For small $\theta$, equation (\ref{eqn_thetaobs_and_theta}) can be simplified:
\begin{equation}
  \theta_{\rm obs} = 2 \theta, 
\end{equation}
which means the relation between emission point and line of sight for a quadrupole field. For a dipole magnetic field, this relation becomes: $\theta_{\rm obs} = 1.5 \theta$ (Tong et al. 2021).
Therefore, the quadrupole field has a wider emission beam than the dipole field.

Using the expression for the $l$-th multipole field (equation (\ref{eqn_multipole})), similar calculations show that the relation between emission point and line of sight is: $\phi_{\rm obs} =\phi$. For the relation between $\theta_{\rm obs}$ and $\theta$ (emission beam width):
\begin{equation}
  \tan\theta_{\rm obs} = \tan\theta \frac{P_l (\cos\theta) + P_l^{\prime}(\cos\theta) \cos\theta /(l+1)}{P_l(\cos\theta) - P_l^{\prime}(\cos\theta) \sin^2\theta/((l+1) \cos\theta)}.
\end{equation}
For a small $\theta$, the properties of Legendre polynomial are: $P_l(1)=1,\ P_l^{\prime}(1)=l(l+1)/2$. Therefore, when $\theta$ is small enough, the relation between emission point and line of sight for a $l$-th order multipole field is:
\begin{equation}
\label{eqn_thetaobs_multipole}
  \theta_{\rm obs} \approx \left( \frac{l}{2} +1 \right) \theta.
\end{equation}
Equation (\ref{eqn_thetaobs_multipole}) means that the opening of the radio emission beam at $(r,\theta)$ is wider in the presence of multipole field.

Traditionally, the pulsar emission beam is assumed to be narrower at higher frequency, i.e. RFM (Radhakrishnan \& Cooke 1969; Ruderman \& Sutherland 1975; Cordes et al. 1978; Wang et al. 2013). However, some pulsars also have a wider emission beam at higher frequency (anti-RFM, Chen \& Wang 2014; Posselt et al. 2021). In our opinion, the presence of multipole field near the neutron star surface may explain the anti-radius-to-frequency mapping behavior:
\begin{enumerate}
   \item At a certain frequency, the radio emission may be the emission at a specified emission height. The emission beam have a width of $1.5\theta$ for a dipole magnetic field.
   \item At higher frequency, the radio emission may emit at lower height. For a dipole field, the emission beam will be narrower at a lower emission height (smaller $\theta$ at the last open field line).
   \item  However, at lower emission height, the strength of the multipole field may dominate over that of the dipole field. In the presence of multipole field, the emission beam may be wider at a lower emission height. Then, a wider pulse profile may be observed at a higher frequency. A simple but quantitative calculation is presented in Section \ref{section_RFM}.
 \end{enumerate}

By using self-similar method, Gourgouliatos (2008) calculated the analytical solution of force-free magnetic field in azimuthal symmetry and arcade topology. It is found that the solved field lines are composed of magnetic arcades, and that these magnetic arcades have angle extent $\Delta \theta$ at NS's surface. These magnetic arcades may be approximated by a multipole field with order $l\sim \pi/\Delta \theta$ (as shown in Fig. 6 of Gourgouliatos 2008). Then the emission beam width for a magnetic arcade is also $(l/2+1) \theta$. An axisymmetric potential multipole magnetic field of order $l$ is a very strong assumption. However, a magnetic arcade may be more possible in NS's magnetosphere, because the magnetic field satisfies $B_{l} \propto \frac{1}{r^{l+2}}$, and then the multipole field decays faster than the dipole field with distance. Then the magnetic arcade may be approximated by a multipole field, and we consider the multipole component exists near NS's surface.

\section{Rotating vector model in the presence of multipole magnetic field}

\begin{figure}
	\centering
	\includegraphics[width=0.5\textwidth]{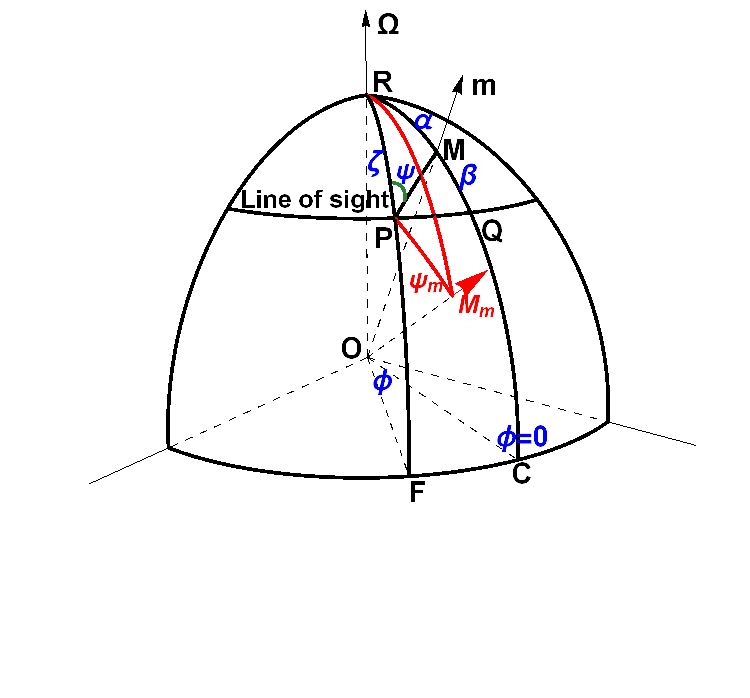}
	\caption{Geometry of the rotating vector model in the presence of multipole field. The vector $\bf m$ is the dipole axis, while the vector $\bf {M_{\rm m}}$ is the multipole axis. The radians $\wideparen{\rm RM}=\alpha$, $\wideparen{\rm MQ}=\beta$, $\wideparen{\rm RP}=\zeta$, and $\wideparen{RPM}=\psi$. At the emission point $\rm P$, the position angle of the rotational axis is different with respect to the dipole axis ($\psi$) and multipole axis ($\psi_{\rm m}$). $\psi_{\rm m}$ may even changes sign with respect to the dipole case. Adapted from Fig. 1 in Tong et al. (2021).}
	\label{fig_grotatingmultipole}
\end{figure}

The position angle as a function of pulse phase can be computed using two different methods: spherical trigonometry and differential geometry, see appendix A and B in Tong et al. (2021). When using spherical trigonometry, the position angle at point $\rm P$ is the angle $\angle \rm{RPM}$ (green arc in Fig. \ref{fig_grotatingmultipole}) in the spherical triangle, formed by three vectors: spin axis, magnetix axis and line of sight. It does not depend on whether the axis is the axis of dipole field or the axis of multipole field. More generally, the RVM only requires a ``rotating vector'', it can be any vector (Manchester 1995). The only requirement is that the projection of magnetic field line on the sphere centred on the neutron star is a great circle (such as a pure dipole or quadrupole field). Usually, the vector is assumed to be the dipole axis. For an axisymmetric potential field, the expression of the polarization position angle for a multipole field is the same as the dipole case:
\begin{equation}\label{eqn_RVM}
  \tan(\psi -\psi_0) = \frac{\sin\alpha \sin(\phi-\phi_0)}{\cos\alpha \sin\zeta -\sin\alpha \cos\zeta \cos(\phi-\phi_0)},
\end{equation}
where $\psi$ is the position angle, $\alpha$ is the angle between the rotational axis and the magnetic axis (which now may be the dipole or multipole field axis), $\zeta$ is the angle between the rotational axis and the line of sight, $\phi$ is the pulse phase, $\phi_0$ and $\psi_0$ are constants during the fitting process, see Fig. \ref{fig_grotatingmultipole} for illustrations. When $\phi=\phi_0$, the position angle is $\psi=\psi_0$. $\psi_0$ is also the position angle of the rotational axis projected onto the plane of sky (Johnston \& Weisberg 2006).

The expression of RVM is unchanged for a multipole field. This point can also be proved using differential geometry.
From Hibschman \& Arons (2001), Tong et al. (2021, appendix B there), the position angle is:
\begin{equation}
  \tan\psi = \frac{\hat{\bf b} \cdot \hat{\Omega}}{\hat{\bf b} \cdot (\hat{\bf t} \times \hat{\Omega})},
\end{equation}
where $\hat{\bf b}$ is the binormal vector at the emission point, $\hat{\bf t}$ is the tangent vector (coincident with the line of sight), $\hat{\Omega}$ is the unit vector along the rotational axis. For a dipole or multipole field, the binormal vector always lies along the toroidal direction: $\hat{\bf b} = \hat{\phi}$. Quantitative calculation for a quadrupole field confirms this point. This is because for an axisymmetric potential field, the magnetic field has no toroidal component. Then the toroidal direction is always a normal vector. Since $\hat{\Omega}$, $\hat{\bf b}$ and $\hat{\bf t}$ are all the same for a dipole or multipole field, the expression for position angle will also be the same for a dipole or multipole field.

The long term evolution of position angle is determined by the large scale dipole field of pulsars and magnetars. At the outburst of magnetars, the catastrophic loss of equilibrium of the flux rope creates a current sheet, which provides an ideal place for magnetic reconnection (Yu \& Huang 2013). The magnetic reconnection process can be approximated by the appearance and disappearance of some mutipole field. Then during the outburst of magnetars, the position angle may be determined by the axis of multipole field. The expression for the position angle is the same, except for a different inclination angle $\alpha$ and $\phi_0$. The other constants, especially $\zeta$ and $\psi_0$ are expected to be the same in the above scenario.

Lower et al. (2021) reported eight observations of the radio emission magnetar Swift J1818.0-1607 over a period of five months. Lower et al. conducted geometric fitting of its linearly polarized position angle by using the RVM. They found that the inclination angle $\alpha$ of Swift J1818.0-1607 changed, and the linearly polarization position angle swing reversed. Compared with the observations 15 days before and 12 days after MJD 59062, the magnetic field geometry changed from $\alpha = 82^\circ$ and $\Psi_{0} = 72.2^\circ$ to $\alpha = 115^\circ$ and $\Psi_{0} = -71^\circ$. Equation (\ref{eqn_RVM}) in this paper can be considered to fit the long term evolution of position angle of Swift J1818.0-1607, and the appearance and disappearance of multipole magnetic field can be used to interpret the changes of magnetic inclination $\alpha$ and polarization position angle of Swift J1818.0-1607. After calculating, the corresponding position angle should be fitted by the RVM of another vector: 
\begin{equation}
  \tan(\psi-\psi_0) = \frac{\sin\alpha_{\rm m} \sin(\phi- \phi_{0,\rm m})}{\cos\alpha_{\rm m} \sin\zeta -\sin\alpha_{\rm m} \cos\zeta \cos(\phi- \phi_{0,\rm m})},
\end{equation}
where $\alpha_{\rm m}$ and $\phi_{0,\rm m}$ is determined by the position of the multipole field axis. It turns out that the position angles of dipole and multipole magnetic field satisfy the same expression, and there is only two constants ($\alpha_{\rm m}$ and $\phi_{0,\rm m}$) that differ between them. $\Delta \alpha =\alpha_{\rm m} -\alpha$ and $\Delta \phi_0 =\phi_{0,\rm m} -\phi_0$ reflects the off-set between the dipole axis and multipole axis.

A different rotating vector can only result in a different $\alpha$ and $\phi_0$. This may be the case in fast radio bursts (FRBs) (Luo et al. 2020). In FRBs, the position angle changes on a shorter timescale compared with that in magnetars. This may reflect a more dynamic magnetosphere of in the case of fast radio bursts. If RVM is employed to fit the position angle of FRBs, different bursts may require different $\alpha$ and $\phi_0$. But the parameters $\zeta$ and $\psi_0$ will not change for different bursts. This is the difference between a dynamic magnetosphere in magnetars (and FRB) and a stable magnetosphere in normal pulsars. 
Simalarly, the position angle swing of magnetar XTE J1810-197 is different from a typical S-like swing (Kramer et al. 2007), which may result from the existence of a multipole field component that can change the magnetosphere structure (i. e. $\alpha$ and $\phi_0$).

\section{Radiation beam evolution in coexistence of dipole and quadrupole fields}
\label{section_RFM}

For simplification, the evolution of radiation beam in a pure dipole magnetic field, a pure quadrupole magnetic field, and a superposition of dipolar and quadrupolar fields are taken as examples in this section. In these different cases, the radius-to-frequency mapping is employed to calculate the relation between radiation beam radius and frequency.
 
 \subsection{Pure dipole field}
\label{dipole_only}

The dipole magnetic axis is assumed to be parallel to the rotational axis. From the previous text, in spherical coordinates, the dipole magnetic field can be expressed as
\begin{equation}
	{\bf{B}}_{\rm dip} = \frac{B_{1}}{r^3}2(\cos\theta \hat{r}+\frac{1}{2}\sin\theta\hat{\theta})
	 \propto (\cos\theta \hat{r}+\dfrac{1}{2}\sin\theta\hat{\theta}).
\end{equation}
For a dipole magnetic field, the tangent vector is expressed as 
\begin{equation}
	\hat{t}_{1} = \frac{1}{N_{1}}(\cos\theta\ ,\frac{1}{2}\sin\theta)
= \frac{1}{N_{1}}(\cos\theta \hat{r}+\frac{1}{2}\sin\theta\hat{\theta}),
\end{equation}
where $N_{1}=\sqrt{1-\frac{3}{4}\sin^2\theta}$ is the normalization constant.
The curvature vector for the field line is (Hibsman \& Arons 2001; Tong et al. 2021): 
\begin{equation}
\vec{\kappa}=(\hat{t}\cdot\nabla)\hat{t} = -\hat{t} \times (\nabla \times \hat{t}).
\end{equation}
The curvature radius is related to the curvature vector as:
\begin{equation}
  \rho = \frac{1}{\lvert{\vec{\kappa}}\rvert} = r \frac{ (5+3\cos2\theta)^{3/2}}{3\sqrt{2}\sin\theta(3+\cos{2\theta})}.
\end{equation}
Using the expression of dipole field mentioned earlier $r=r_{\rm e} \sin^{2}\theta$, the curvature radius becomes:
\begin{equation}\label{rho_theta_re-dipole}
	\rho=r_{\rm e}\frac{\sin\theta(5+3\cos2\theta)^{3/2}}{3\sqrt{2}(3+\cos{2\theta})}.
\end{equation}
When $\theta$ is a small angle, we can use estimations $\sin\theta \approx \theta$ and $\cos\theta \approx 1$. Then equation (\ref{rho_theta_re-dipole}) can be estimated to be
\begin{equation}\label{rho_theta_dipole}
	\rho \approx \frac{4}{3} r_{\rm e} \theta.
\end{equation}
Assuming that the emission beam of pulsar radio emission is determined by the opening angle of the last closed field line (see Section \ref{section_Relation} for details), the radiation beam radius is figured to be
\begin{equation}\label{dipole:rhobeam_theta_rho}
	\rho_{\rm beam} =\theta_{\rm obs} = 1.5\theta=\frac{9}{8}\frac{\rho}{r_{\rm e}}.
\end{equation}
If a dipole magnetic field is assumed to produce curvature radiation, it can produce radio radiation of frequency  $\nu=\frac{3\gamma^{3}c}{4\pi\rho}$ (Wang et al. 2013). After plugging it into the above equation, the beam radius is
\begin{equation}\label{radiation radius of a dipole magnetic field}
	\rho_{\rm beam}=\frac{27}{32\pi} \frac{\gamma^{3}c}{r_{\rm e}\nu}.
\end{equation}
For simplification, we consider $r_{\rm e} \approx r_{\rm LC}=\frac{c  P}{2\pi}$. The focus of this paper is to highlight the physics and model construction of pulsar magnetic field and radio radiation, and mainly discuss the dependence between radiation beam and frequency. Here we make the assumption that the Lorentz factor $\gamma$ is a constant, which is a simplification of the calculation process. For instance, we use the data in Table 1 of Gangadhara (2004) to test our model. When frequency $\nu=325$ MHz, Lorentz factor $\gamma=286$, we use the formula (\ref{radiation radius of a dipole magnetic field}) in our paper to calculate and convert the result into the unit of degree. Then the radiation beam width is $\rho_{\rm beam}\approx7.0^{\circ}$ in this case, which is consistent with the observations.

Later we combine equation (\ref{radiation radius of a dipole magnetic field}) with $r=r_{\rm e} \sin^{2}\theta \approx r_{\rm e}\theta^{2}$ (for a small $\theta$) and frequency of curvature radiation $\nu=\frac{3\gamma^{3}c}{4 \pi \rho}$. The dipole field dominates far from the pulsar, and the Lorentz factor of the particle here satisfies 
\begin{equation}
	\gamma_{\rm d}=(\frac{16\pi \nu \sqrt{r\cdot r_{\rm e}}}{9c})^{\frac{1}{3}}.
\end{equation}
It can be estimated from Fig. 5 of Shang et al. (2017) that $r \approx 155$ km is where the radiation with a frequency of $2$ GHz is generated. Using the data from PSR B0329+54 mentioned above, we find that the Lorentz factors are $\gamma_{\rm d} \approx 441$ (this is consistent with the observation).

From equation (\ref{radiation radius of a dipole magnetic field}), the radiation beam radius of dipole magnetic field is inversely proportional to the frequency in the pure dipole case, that is $\rho_{\rm beam} \varpropto \nu^{-1}$. This result is consistent with the observation of some pulsars of which the radio profile gradually becomes narrower as frequency increases. However, it cannot work when the radio radiation profile of pulsar is basically unchangeable or gradually widens with the increasing frequency (Chen \& Wang 2014; Posselt es al. 2021; Agar et al. 2021). Hence it is then assumed that the multipole magnetic field and the dipole magnetic field coexist in the magnetosphere of pulsars. The quadrupole field is taken as an example.

\subsection{Pure quadrupole field}
\label{quadrupole_only}

The calculation for a pure quadrupole field is similar to the above dipole case. In spherical coordinates, pure quadrupolar field is given by 
${\bf B}_{\rm quad} \varpropto (\frac{3\cos^{2}\theta-1}{2},\sin\theta \cos\theta,0) \varpropto (\frac{3\cos^{2}\theta-1}{2} \hat{r}+\sin\theta \cos\theta \hat{\theta})$ 
(see Section \ref{section_multipole magnetic field}). Previously, the radiation angle and profile width are calculated from the last open field line of the dipole magnetic field. Because the key point of this section is to obtain the variation of magnetic field lines and beam widths at the same $(r,\theta)$ before and after applying a quadrupole field, it is assumed that the relationship between $r$ and $r_{\rm e}$ is still $r=r_{\rm e} \sin^{2}\theta$ to calculate the evolution of $\rho_{\rm beam}$. 
The curvature radius in quadrupole case is calculated to be:
\begin{equation}
	\rho=r\frac{[12\cos{2\theta}+5(3+\cos{4\theta})]^{3/2}\csc{2\theta}}{2\sqrt{2}(39+20\cos{2\theta}+5\cos{4\theta})};
\end{equation}
substituting $r=r_{\rm e} \sin^{2}\theta$ into the above formula, we obtain:
\begin{equation}
	\rho=r_{\rm e}\frac{[12\cos{2\theta}+5(3+\cos{4\theta})]^{3/2}\csc{2\theta}\sin^2{2\theta}}{2\sqrt{2}(39+20\cos{2\theta}+5\cos{4\theta})}.
\end{equation}
For a small $\theta$, the curvature radius of quadrupole magnetic field can be calculated as
\begin{equation}
	\rho=\frac{r_{\rm e}}{2}\theta.
\end{equation}
Finally, the radiation beam radius of quadrupole field is figured out to be 
\begin{equation}\label{quadrupole:rhobeam_theta_rho}
	\rho_{\rm beam}=2\theta=4\frac{\rho}{r_{\rm e}}.
\end{equation}
Plug the relation due to curvature radiation $\nu=\frac{3\gamma^{3}c}{4\pi\rho}$ (Wang et al. 2013) into the formula above, then the radiation radius of quadrupole field has a relation to frequency $\nu$, by
\begin{equation}\label{radiation radius of a quadrupole magnetic field}
	\rho_{\rm beam}=\frac{3}{\pi}\frac{\gamma^{3}c}{r_{\rm e}\nu}.
\end{equation}
Consistent with the expectations in Section \ref{section_Relation}, the beam radius is indeed wider than that in the dipole case (see eq.(\ref{radiation radius of a dipole magnetic field})). However, for a pure quadrupole field, the beam radius is also narrower at higher frequency ($\rho_{\rm beam} \propto 1/\nu$). As a result, the coexistence of dipole and quadrupole magnetic field should be considered.

We adopt the same processing method as in the previous section, and first assume the curvature radius as $\rho \approx r_{\rm NS}$ in this case, thus the Lorentz factor in quadrupolar case satisfies:
\begin{equation}
	\gamma_{\rm q} \approx (\frac{4\pi\nu r_{\rm NS}}{3c})^\frac{1}{3}.
\end{equation}
This is because the quadrupole field decays faster with distance $r$ than dipole field, and then dominates around the neutron star. Using the data from PSR B0329+54 mentioned above, we find that the Lorentz factor is $\gamma_{\rm q} \approx 986$, which is consistent with the observations.

\subsection{An aligned dipole and quadrupole field}
\label{dipole and quadrupole, aligned}

In the case of which the dipole and quadrupole magnetic field coexist and their magnetic axes are aligned, $b_{\rm q}$ is supposed to be the magnitude ratio of quadrupole and dipole field at the magnetic pole on pulsar's surface. In order to unify the calculation, the constants $B_{1}$ and $B_{2}$ in the pure multipole fields are transformed into the surface magnetic field at the magnetic pole. These two magnetic field structures satisfy $B_{\rm dip,p}=\frac{2B_{1}}{R^3}$ and $B_{\rm quad,p}=\frac{3B_{2}}{R^4}$ respectively at magnetic pole. Then their potential can be written through magnetic field as
\begin{eqnarray}
	\phi_{\rm dip} &=& \frac{B_{\rm dip}}{2}\frac{R^{3}}{r^2}\cos{\theta} \quad (l=1) \\
	\phi_{\rm quad} &=& \frac{B_{\rm quad}}{3}\frac{R^{4}}{r^3}\frac{3\cos^{2}\theta-1}{2} \quad (l=2)
\end{eqnarray}
where $R$ stands for the neutron star's radius. It can be written likewise that 
\begin{equation}
	\phi_{l}=\frac{B_{l,pole}}{l+1}\frac{R^{l+2}}{r^{l+1}}P_{l}(\cos\theta)
\end{equation} 
for multipole field.

Hence the potential can be written as 
\begin{equation}
	\phi_{\rm{tot}} = \frac{B_{\rm dip}}{2}\frac{R^3}{r^2}\cos{\theta} + \frac{B_{\rm quad}}{3}\frac{R^4}{r^3}\frac{3\cos^{2}\theta-1}{2}
\end{equation}
Here we make a dimensionless treatment for the rardation height $r$ and the magnetic field strength $B$, and assume that $r$ is in units of the neutron star's radius $R$, $B_{\rm dip}=1$ and $B_{\rm quad}=b_{\rm q}$ in this paper ($B_{\rm dip}$ is taken as the unit value, and $b_{\rm q}$ means the relative strength  of quadrupole and dipole field). After the same calculation as before, the magnetic field in the aligned case is 
\begin{equation}
	{\bf B}_{\rm a} = (\frac{\cos\theta}{r^3} + b_{\rm q}\frac{3\cos^{2}\theta-1}{2r^4})\hat{r} + (\frac{\sin\theta}{2r^3} + b_{\rm q}\frac{\cos{\theta} \sin{\theta}}{r^4})\hat{\theta}.
\end{equation}
We just consider a simplified treatment of pulsar’s magnetosphere, that is, the dipole and quadrupole fields with the coincided center. Then the expression of the last open magnetic field line will change. But in this paper, the main focus here is the effect of whether the quadrupole field exists at the same $(r,\theta)$ on the field lines and beam width. Therefore, the last open field line is  $r=r_{\rm e} \sin^{2}\theta$. After estimation, the curvature radius is 
\begin{equation}
	\rho=\frac{r_{\rm e}\theta}{2}.
\end{equation}

The opening angle of magnetic field line (i.e. the angular radius of radiation beam $\rho_{\rm beam}$) is defined by $\cos{\rho_{\rm beam}}=\hat{z} \cdot \hat{t}$ ( $\hat{t}$ is the tangent vector of the whole magnetic field, and $\hat{z}=\cos{\theta}\hat{r}-\sin{\theta}\hat{\theta}$ means a unit vector along the magnetic moment). For a small $\theta$, $r=r_{\rm e} \sin^2\theta$ is also very small. Therefore, the curvature is dominated by quadrupole field. This may explain why the curvature radius is the same as the pure quadrupole case.
Meanwhile, for a small $\theta$, the relationship bewtween $\rho_{\rm beam}$ and $\theta$ is 
\begin{equation}\label{rho_beam dipole+quadrupole 1}
	\rho_{\rm beam}=\frac{4b_{\rm q}+3r}{2(b_{\rm q}+r)}\theta
\end{equation}
The result of equation (\ref{rho_beam dipole+quadrupole 1}) describes the pure dipole case when $b_{\rm q}=0$, and if $r \gg b_{\rm q}$, the magnetic field is dominated by dipole field, and $\rho_{\rm beam} \approx \frac{3}{2}\theta$; on the contrary, it approximates the pure quadrupole case when $b_{\rm q}$ approaches infinity, and if $r \ll b_{\rm q}$, the magnetic field is dominated by quadrupole field, and the radiation radius $\rho_{\rm beam} \approx 2\theta$. It can be seem that equation (\ref{rho_beam dipole+quadrupole 1}), which portrays the case about aligned dipole and quadrupole field, is consistent with the results of pure dipole field and pure quadrupole field in the previous sections in some cases.

To obtain the trend of $\rho_{\rm beam}$ with frequency $\nu$, the relationship $r=r_{\rm e} \sin^{2}\theta$ can be converted to $\theta \approx \sqrt{\frac{r}{r_{\rm e}}}$ for a small $\theta$, and then $r=\frac{4\rho^2}{r_{\rm e}}$ can be get by using the expression of curvature radius $\rho \approx \frac{r_{\rm e}}{2}\theta \approx \frac{\sqrt{r \cdot r_{\rm e}}}{2}$. For more quantitative calculation, the formulae (\ref{rho_beam dipole+quadrupole 1}), $\rho \approx \frac{r_{\rm e}}{2}\theta$ and $\theta \approx \frac{2\rho}{r_{\rm e}}$ are combined to rewrite the expression of beam radius: 
\begin{eqnarray}
	\rho_{\rm beam} &=& \frac{2(b_{\rm q}+3\frac{\rho^2}{r_{\rm e}})}{b_{\rm q}+\frac{4\rho^2}{r_{\rm e}}}\frac{2\rho}{r_{\rm e}} \\
	&=& \frac{4b_{\rm q}}{b_{\rm q}+4\frac{\rho^2}{r_{\rm e}}}\frac{\rho}{r_{\rm e}}+\frac{12\frac{\rho^2}{r_{\rm e}}}{b_{\rm q}+4\frac{\rho^2}{r_{\rm e}}}\frac{\rho}{r_{\rm e}}.
\end{eqnarray}
A variable $\rho_{\rm 0}$ is assumed to be $\rho_{\rm 0}({\nu})=\frac{\rho}{r_{\rm e}}=\frac{1}{r_{\rm e}}\frac{3\gamma^3c}{4\pi }\frac{1}{\nu} \propto \frac{1}{\nu}$ (for a pure quadrupole or a dipole field case). Finally, $\rho_{\rm beam}$ is found to be 
\begin{equation}
	\rho_{\rm beam}(\nu)=\frac{4b_{\rm q}}{b_{\rm q}+4\rho_{\rm 0}^2r_{\rm e}}\rho_{\rm 0}+\frac{12\rho_{\rm 0}^2r_{\rm e}}{b_{\rm q}+4\rho_{\rm 0}^2r_{\rm e}}\rho_{\rm 0}.
\end{equation}
If the period of a pulsar is given, the typical radius of radiation beam $\rho_{\rm 0} \propto \frac{1}{\nu}$. When quadrupole field and dipole field are present simultaneously, the radiation beam becomes a complex function of $\nu$. If the ratio of quadrupole field to dipole field is known, then the frequency dependence of radiation beam can be obtained.

Next, the radiation beam radius is processed dimensionlessly. It is assumed that $\nu=\nu_{\rm eq}$ and $\theta=\theta_{\rm eq}$ at distance $r=b_{\rm q}$ where the magnitude of these multipole fields is equal, then we can get $r=\frac{4\rho^2}{r_{\rm e}}=b_{\rm q}(\frac{\nu_{\rm eq}}{\nu})^2 \propto \frac{1}{\nu^2}$, $\theta=\frac{2\rho}{r_{\rm e}}=\theta_{\rm eq}\frac{\nu_{\rm eq}}{\nu} \approx \frac{1}{\nu}$, where $\theta_{\rm eq}$ is the corresponding polar angle at the radius $r=b_{\rm q}$. Suppose $\nu'=\frac{\nu}{\nu_{\rm eq}}$. After that, equation (\ref{rho_beam dipole+quadrupole 1}) becomes 
\begin{equation}
	\rho_{\rm beam}=\frac{4+3\frac{1}{\nu'^2}}{2(1+\frac{1}{\nu'^2})}\theta_{\rm eq}\frac{1}{\nu'}.
\end{equation}
Introducing the dimensionless beam radius: $\rho_{\rm beam}'=\rho_{\rm beam}/\theta_{\rm eq}$, and then the beam radius can be written as:
\begin{equation}\label{dimensionless beam radius}
	\rho'_{\rm beam}=\frac{3}{2(1+\nu'^2)}\frac{1}{\nu'}+\frac{2\nu'^2}{1+\nu'^2}\frac{1}{\nu'}.
\end{equation}
The first and second terms on the right side of equation (\ref{dimensionless beam radius}) represent dipole and quadrupole components. It is found that the frequency dependence of $\rho'_{\rm beam}$ is complicated, but anti-radius-to-frequency mapping cannot be given from this relation, probably because $\theta \ll 1$ and $r \ll r_{\rm e}$ are set in the calculation process, which makes the quadrupole component play the leading role. In equation (\ref{dimensionless beam radius}), the relation $\rho_{\rm beam} \propto \frac{1}{\nu}$ dominates, but $\rho_{\rm beam}$ can have a more complex frequency dependence.

It is further speculated that $\rho_{\rm beam}$ is mainly determined by radius in the region where the quadrupole field is roughly equivalent to the dipole field (it follows from the expression of these multipole fields that such a region is not large, so $\theta$ can be regarded as a constant):
\begin{eqnarray}
	\rho_{\rm beam} &=& \frac{4b_{\rm q}+3r}{2(b_{\rm q}+r)}\theta \\ 
	&=& \frac{3}{2}\theta+\frac{b_{\rm q}}{2(b_{\rm q}+r)}\theta.
\end{eqnarray}
After introducing the dimensionless radiation beam radius $\rho'_{\rm beam}$, formula (\ref{rho_beam dipole+quadrupole 1}) translates to 
\begin{eqnarray}
	\rho'_{\rm beam} &=& \frac{3}{2}+\frac{1}{2(1+\frac{1}{\nu'^{2}})} \\ 
	&=& \frac{3}{2}+\frac{\nu'^2}{2(1+\nu'^{2})} \label{rho'_beam1} ,
\end{eqnarray} 
which has the same form as that used in Chen \& Wang (2014) (that is, has constant term and frequency dependence). 
According to the equation (1) in Chen \& Wang et al. (2014), the radiation beam radius $\rho_{\rm beam}$ and frequency $\nu$ satisfy $\rho_{\rm beam} =\rho_{\rm 0}+\rho_{\rm c} {\nu}^{\mu}$, where $\mu$ is the power index reflecting the evolving tend.
\begin{figure}
	\centering
	\includegraphics[width=0.45\textwidth]{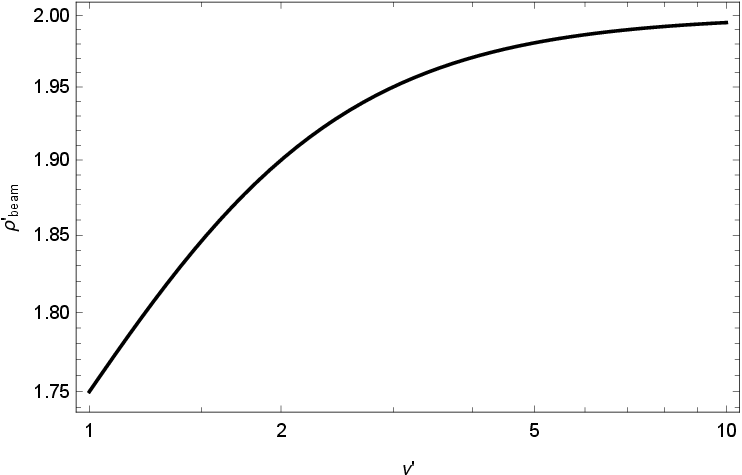}
	\caption{Diagram depicting the trend between the dimensionless radiation beam radius and frequancy.}
	\label{dimensionless pho_beam-nu}
\end{figure}

In a certain frequency range, $\rho_{\rm beam}$ trend of pulsars with parallel dipole and quadrupole fields is shown in Fig. \ref{dimensionless pho_beam-nu}. So it turns out that $\rho_{\rm beam}$ widens with increasing frequency.
Then RFM in the presence of multipole field can be considered as a geometric constraint on the magnetosphere of pulsars to explain the radio profile evolution of pulsars, such as the pulse broadening samples in Chen \& Wang (2014). By calculating the relative fraction of pulse width change between 0.4 GHz and 4.85 GHz, the result in Chen \& Wang (2014) shows that the profiles of $46\%$ of the samples widen at high frequencies, including the whole group-C ($\eta > 10\%$) pulsars and a portion of group-B ($-10\% \le \eta \le 10\%$) pulsars with slight pulse broadening.
The pulsars having the magnetosphere model described in this paper belong to group C in Chen \& Wang (2014). Thus the model in this paper can interpret the observed anti-radius-to-frequency mapping phenomena.

Some more anti-radius-to-frequency mapping phenomena have been found recently. Posselt et al. (2021) used MeerKAT telescope to observe a sample of radio pulsars. Through defining width color, $C_{\bf xy}$, and width contrasts, $K_{\bf xy}$, Posselt et al. discovered that from the 420 pulsars observed with effective pairwise ratios, about one-third of the pulsars' profiles widen with increasing frequency. Besides, through broadband radio observation of the slowest-spinning radio pulsar PSR J0250+5854, Agar et al. (2021) found that $W_{50}$, the pulse width at $50\%$ of the peak value, of PSR J0250+5854 increases towards higher frequency, from around 1\textdegree at 150 MHz to 2\textdegree at 1250 MHz. These observations are difficult to be explained or predicted using conventional radius-to-frequency mapping. But in our model, an explanation for these observations becomes possible.

\subsection{A magnetic field with a misaligned dipole and quadrupole field}
\label{dipole and quadrupole, misaligned}

\begin{figure}
	\centering
	\includegraphics[width=0.45\textwidth]{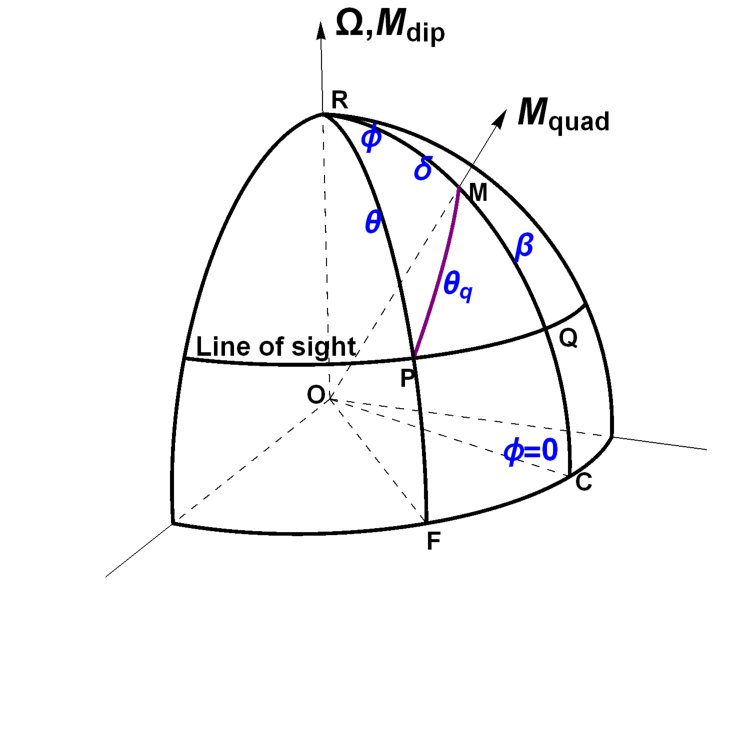}
	\caption{The magnetic field structure with misaligned quadrupole and dipole field. The magnetic axis of dipole field coincides with the rotation axis of pulsar. P is a point located in line of sight at a certain time, and its angles between quadrupole field and dipole field are $\theta_{\rm q}$ and $\theta$ respectively. The magnetic inclination angle of quadrupole field is $\delta$.}
	\label{misaligned dip + quad}
\end{figure}

Our present calculation assumes that $\theta$ is a small quantity and also that the angle between the magnetic axis of dipole and quadrupole field is a small quantity. Based on these assumptions, we can obtain the analytical solution of the magnetic field expression. General results may rely on numerical calculations.

From formula (\ref{rho_beam dipole+quadrupole 1}), it can be imagined that the radiation beam radius is proportional to $\theta$ at point $(r,\theta)$, as shown in the case of pure multipole field above; and the radius dependence, the previous term, leads to a richer frequency dependence. Similarly, when quadrupole and dipole field coexist with non-aligned magnetic axes, $\rho_{\rm beam}$ can be assumed to satisfy $\rho_{\rm beam}=f(r,\phi)\theta$, where $f(r,\phi)$ is some function to reflect the dependence of $\rho_{\rm beam}$ on $r$ and $\phi$.

The magnetosphere with non-aligned magnetic axes is shown in the Fig. \ref{misaligned dip + quad}. For simplification, the magnetic axis of dipole field is set to coincide with the spin axis of the pulsar. Assuming that the angle between magnetic axis of two multipole fields and point P are $\theta$ and $\theta_{\rm q}$ respectively. From spherical geometry $\cos{\theta_{\rm q}}=\cos{\theta}  \cos{\delta}+\sin{\theta} \sin{\delta} \cos{\phi}$, the quadrupole potential can be expressed as
\begin{equation}
	\phi_{\rm quad}=\frac{B_{\rm quad}}{3}\frac{R^{4}}{r^3}\frac{3\cos^{2}\theta_{\rm q}-1}{2},
\end{equation}
and potential with the coexistence of dipole and quadrupole field is
\begin{equation}
	\phi_{\rm tot}=\frac{B_{\rm dip}}{2}\frac{R^{3}}{r^2}\cos{\theta}+\frac{B_{\rm quad}}{3}\frac{R^{4}}{r^3}\frac{3\cos^{2}\theta_{\rm q}-1}{2}.
\end{equation}
By using ${\bf B}=-\nabla \phi_{\rm tot}$ ($\nabla=\hat{r}\frac{\partial}{\partial r}+\hat{\theta}\frac{1}{r}\frac{\partial}{\partial \theta}+\hat{\phi}\frac{1}{r\sin\theta}\frac{\partial}{\partial \phi}$), we can get the expression of magnetosphere. It follows that $\phi_{\rm tot}$ is also dependent on $\phi$ , which suggests $\bf{B}$ has $\hat{r}$, $\hat{\theta}$ and $\hat{\phi}$ compoenents. But the angles $\delta$, $\theta$ and $\theta_{\rm q}$ can be small enough, while the angle $\phi$ cannot just stay small, as it evolves over time.

Use the same calculation process as in the previous section. When $\theta$ and $\delta$ are small, beam radius is
\begin{equation}\label{rho_beam(r,bq,phi), mis q+d}
	\rho_{\rm beam}=\frac{(4b_{\rm q}+3r)\theta-2b_{\rm q}\delta\cos{\phi}}{2(b_{\rm q}+r)}.
\end{equation}
In this equation, with the growth of $\delta$ or $b_{\rm q}$, $\rho_{\rm beam}$ deviates more and more from that in the case when the quadrupolar and dipoar coincide. It can be obviously seen that equation (\ref{rho_beam(r,bq,phi), mis q+d}) will go back to the previous case of an aligned dipole and quadrupole field if $\delta = 0$. As for the quantity $\phi$ that represents the phase, $\rho_{\rm beam}$ has a minimum value when $\phi=$ 0\textdegree; $\rho_{\rm beam}$ becomes larger as $\phi$ increase, and it arrives at the maximum value when $\phi=$ 180\textdegree. The shape of the radiation beam is no longer a circular. When the axis of the dipole and quadrupole have a off-set, the frequency dependency of $\rho_{\rm beam}$ could be obtained with the method in Subsection \ref{dipole and quadrupole, aligned}. Then the equation (\ref{rho_beam(r,bq,phi), mis q+d}) is transformed into 
\begin{equation}
	\rho'_{\rm beam}=\frac{3}{2}+\frac{\nu'^{2}(1-\frac{2\delta \cos{\phi}}{\theta})}{2(1+\nu'^{2})},
\end{equation}
which has an extra factor $(1-\frac{2\delta \cos{\phi}}{\theta})$ compared with equation (\ref{rho'_beam1}). Whether this factor is positive or negative can affect the variation of $\rho'_{\rm beam}$ with $\nu'$, so that the frequency dependence of pulse width conforms radius-to-frequency mapping or anti-radius-to-frequency mapping respectively.

\section{Discussion}

This paper provides an interpretation of RFM and anti-RFM phenomena. In terms of RFM, the radio radiation model of pulsars is divided into narrowband and broadband models. In radio band, only coherent emission can produce enough brightness temperature to be observed, so the radio radiation of pulsars must be coherent (Goldreich \& Keeley 1971). Buschauer \& Benford (1976) proved that the radiation is narrowband if plasma wave causes coherent amplification through its particle density fluctuation. If radio radiation with a given frequency is generated at a particular emission height, the higher the frequency is, the lower emission height becomes. In this paper, narrowband radiation is preferred, and dipole and multipole field are combined to describe the magnetosphere structure and explain RFM and anti-RFM.

As for broadband emission model, a radiation with a broadband frequency can arise from a narrow range in radius, which will lead to observations that do not match RFM. Compared with narrowband emission, broadband emission is more suitable for explaining the range of spectra, pulse shapes, frequencies, and the absence of aberration effects (Buschauer \& Benford 1980). But only narrowband model is considered in this paper to simplify the problem.

During the preparation process of this work, we noted the work of Yamasaki et al. (2022). They adopted a method similar to this paper, considering quadrudipolar magnetic field (including inclined dipolar and quadrupolar magnetic fields) in vacuum. Through adjusting two variables, inclination angle $i_{\rm QD}$ and field strength ratio $f_{\rm Q}$ between dipole and quadrupole magnetic field, they described the shape of multipole magnetic field line and its radio beam. It is also shown that the angular range and radio beam width in quadrudipolar are wider than those in pure dipolar. Unlike Yamasaki et al. (2022), we consider the relationship between line of sight and emission point in the case of the general multipole field in Section 2 and 3, and then calculate the contribution of dipole and quadrupole fields to the beam width by combining them. We use the RVM as geometric constraint. When the quadrupole and dipole field are not aligned, the total magnetic field in our magnetosphere model contains not only r and $\theta$ components, but also $\phi$ component, and we get the tendency of radiation beam radius with frequency later.

In our work, the emission height can be deduced when the specific radiation and Lorentz factor $\gamma$ are determined. In the case of a pure dipole or quadrupole field, the relationship between beam radius $\rho_{\rm beam}$, polar angle $\theta$ and curvature radius $\rho$ are respectively equation (\ref{dipole:rhobeam_theta_rho}) and (\ref{quadrupole:rhobeam_theta_rho}) in this paper. The last opening magnetic field line is supposed and estimated to be $r=r_{\rm e}\sin^{2}\theta \approx r_{\rm e}\theta^{2}$ for a small $\theta$. The curvature radiation is assumed in our paper, and it will produce radio emission of frequency $\nu=\frac{3\gamma^{3}c}{4\pi\rho}$. Combining these formulas, the emission height can be written as
\begin{equation}
	r \approx \frac{81\gamma^{6}c^{2}}{256\pi^{2}r_{\rm e}\nu^{2}} \propto \frac{1}{\nu^{2}}
\end{equation} 
and
\begin{equation}
	r \approx \frac{9\gamma^{6}c^{2}}{4\pi^{2}r_{\rm e}\nu^{2}} \propto \frac{1}{\nu^{2}}.
\end{equation}
for pure dipolar and quadrupolar respectively. Here, the speed of light is $c=3 \times 10^{10} \, \rm {cm/s}$, and $r_{\rm e} \approx r_{\rm LC} =4777 \, r_{\rm NS}$ ($r_{\rm LC}$ means the radius of light speed cylinder). We assume the period and the radius of pusar as $P=1 \, \rm s$ and $r_{\rm NS}=10 \, \rm km=10^{6} \, \rm cm$, and use the parameters, $\nu=800 \, \rm MHz$ and $\gamma=350$, from Wang et al. (2013). As a consequence, the emission heights are $r \approx 1.7 \times 10^{7}\rm \, cm= 17\, \emph{r}_{\rm NS}$ for a pure dipolar, which is in line with the calculation results in Wang et al. (2013), and $r \approx 1.2 \times 10^{8} \, \rm cm =120 \, \emph{r}_{\rm NS}$ for a pure quadrupolar.
This proves the fitness of our model to explain anti-RFM to some extent.

\section{Conclusion}

Here we consider the modification of the rotating vector model and radius-to-frequency mapping for an axisymmetric potential field. There are some assumptions during this process.
\begin{enumerate}
  \item An axissymmetric potential field has no toroidal component. A non-potential field will have a toroidal component, e.g. a force-free field that has a toroidal component (Wolfson 1995, Thompson et al. 2002, Tong 2019). The toroidal field will also modify the RVM. Equation (30) in Tong et al. (2021) presents an estimation for the effect of the toroidal field.
  \item The assumption of axisymmetric is only for the sake of simplicity. The general case will have complicated magnetic field geometry, even for a potential field (Wiegelmann \& Sakurai 2012). For a general field geometry, the corresponding RVM may be calculated using differential geometry (Hibschman \& Arons 2001; Tong et al. 2021).
  \item We mainly consider a single $l$-th multipole, that is when the field is a dipole, or quadrupole, or octupole etc. The actual case may be a mixture of different multipoles, e.g. a dipole field ($10^{12} \ \rm G$ at the neutron star surface) with a quadrupole field ($10^{14} \ \rm G$ at the star surface). Generally, the axis of the dipole and quadrupole have some off-sets. A small angle between the magnetic axis of dipole field and multipole field is considered in this paper. We calculated the geometry of magnetic field and the evolution of radio radiation beam in this case, and found that the beam radius can be described by the formula (\ref{rho_beam(r,bq,phi), mis q+d}). The frequency dependence of radio radiation pulse width is influenced by the positivity and negativity of the factor $(1-\frac{2\delta \cos{\phi}}{\theta})$, which can explain the phenomena of anti-RFM to some extent.
  \item We discuss two observations which may be led to the effect of multipole field: (1) For normal pulsars, the anti-RFM may result from the presence of multipole field at lower emission height. (2) For magnetars (and FRBs), the changing slope of position angle may cause due to the appearance and disappearance of multipole field. There may be various kinds of multipole field in pulsars and magnetars (Thompson et al. 2002; Pavan et al. 2009; Beloborodov 2009; Tong 2019; Bilous et al. 2019). Irrespective of the field geometry, we expect that the corresponding RVM will be modified to some degree.
\end{enumerate}

In summary, the RVM and RFM will be modified in the presence of multipole field. (1) The radiation beam radius decreases with increasing frequency for a pure dipole or quadrupole magnetic field, which can explain RFM. (2) The radio emission beam radius will be larger in the presence of multipole field. (3) When dipole and quadrupole magnetic field coexist, the radiation beam radius follows the form $\rho_{\rm beam}=\rm const + \nu^{\mu}$ roughly in the place where both of multipole fields are roughly equal, and widens with increasing frequency in some frequency range, which can explain anti-RFM. The specific relationship depends on magnitude ratio of these two magnetic components $b_{\rm q}$, magnetic axis angle $\delta$, and phase angle $\phi$. (4) The expression of position angle will be the same, with possible changes of the inclination angle ($\alpha$) and the phase constant ($\phi_0$) parameters. The other constants (such as $\zeta$ and $\psi_0$) will be the same. When fitting the corresponding position angle in magnetars (and FRBs), these constraints should be considered.

\begin{acknowledgements}
H.Tong is supported by National SKA Program of China (2020SKA0120300) and NSFC (11773008). 
\end{acknowledgements}











\end{document}